\documentclass[secnumarabic,amssymb, nobibnotes, aps, prd, preprint,superscriptaddress]{revtex4-1}
\pdfoutput=1

\usepackage{graphicx}
\usepackage{amsmath}
\usepackage{color}
\usepackage{hyperref}

\begin{document}

\title{Asymmetric capture of Dirac dark matter by the Sun}

\author{Mattias Blennow}%
\email[Electronic address: ]{emb@kth.se}
\affiliation{Department  of  Theoretical  Physics,
School  of  Engineering  Sciences, KTH Royal Institute of Technology, Albanova University Center, 106 91, Stockholm, Sweden}
\author{Stefan Clementz}%
\email[Electronic address: ]{scl@kth.se}
\affiliation{Department  of  Theoretical  Physics,
School  of  Engineering  Sciences, KTH Royal Institute of Technology, Albanova University Center, 106 91, Stockholm, Sweden}
\begin{abstract}
Current problems with the solar model may be alleviated if a significant amount of dark matter from the galactic halo is captured in the Sun. We discuss the capture process in the case where the dark matter is a Dirac fermion and the background halo consists of equal amounts of dark matter and anti-dark matter. By considering the case where dark matter and anti-dark matter have different cross sections on solar nuclei as well as the case where the capture process is considered to be a Poisson process, we find that a significant asymmetry between the captured dark particles and anti-particles is possible even for an annihilation cross section in the range expected for thermal relic dark matter. Since the captured number of particles are competitive with asymmetric dark matter models in a large range of parameter space, one may expect solar physics to be altered by the capture of Dirac dark matter. It is thus possible that solutions to the solar composition problem may be searched for in these type of models.
\end{abstract}
\maketitle

\section{Introduction}\label{intro}
Over the past few decades, a large amount of evidence in support of the existence of dark matter (DM) has been assembled, see e.g., refs~\cite{Bergstrom:2000pn,Bertone:2004pz}. To describe the properties of DM has become one of the main issues not only in cosmology but also in particle physics, since the Standard Model leaves no room for its existence, which means extending it is necessary. Some of the most studied particle candidates of DM are weakly interacting massive particles such as the lightest neutral supersymmetric particle, the neutralino (an in-depth review can be found in ref~\cite{Jungman1996}).

If DM interacts with regular matter, it may scatter in astrophysical bodies such as the Sun and become gravitationally bound. Over time, this can lead to a large accumulation of DM, the effect of which used to be one of the proposed explanations of the solar neutrino problem~\cite{Press1985}. This idea was later discarded in favour of neutrino flavour conversion by the neutral current phase of the SNO experiment~\cite{Ahmad2002}. Since the downward revision of heavy elements in the standard solar model \cite{Asplund:2004eu}, theoretical predictions and observations of helioseismology do not match and the Sun now faces the solar composition problem \cite{Asplund2009}. Again, a proposed solution is DM trapped in the Sun due to its interaction with regular matter. The idea is that DM particles collide with nuclei in the Sun, lose enough energy to become gravitationally bound and to eventually settle in the solar core after additional scattering. Once in the core, DM can scatter off of the thermal distribution of nuclei and gain some energy which is then lost by scattering in the outer regions of the Sun, the result of which is a lower core temperature~\cite{Gould1990}. This would ultimately affect helioseismology and possibly provide a cure for the solar composition problem. The shift in temperature of the solar core would also change the solar neutrino fluxes that are observed by experiments, in particular the $^8$B flux which varies as $T^{25}$~\cite{Bahcall1996}. The effects of DM captured by the Sun has been extensively studied for various DM models~\cite{Spergel1985,Dearborn1991,Frandsen2010,Taoso2010,Cumberbatch2010,Lopes2012,Lopes2014} and also in other stars~\cite{Kouvaris2008,Kouvaris2010,Kouvaris2011,Mcdermott2012,Iocco2012}. Annihilating DM can also provide a channel for indirect detection of DM as a flux of high energy neutrinos from these annihilations might be detectable on Earth. Such signals are being searched for by neutrino telescopes~\cite{IceCube2013,Aantares2013}. In the case where the DM has self-interactions, DM particles already captured by the Sun provide a possibility of DM self-capture, which could lead to higher concentrations of DM inside the Sun. If DM annihilations produce high-energy neutrinos, this would increase the number of annihilations and thus the signal in neutrino telescopes could be enhanced~\cite{Zentner2009}. Energy injection into the Sun by the DM annihilation products has been studied, but requires the halo density to be many orders of magnitude higher than the local density for the star to be affected~\cite{Fairbairn2008}.

The models which achieve the highest numbers of captured DM particles in the Sun are models in which annihilation is suppressed or does not occur at all, the latter of which is the case in asymmetric DM (ADM) models, reviews of which can be found in~\cite{Zurek:2013wia,Petraki:2013wwa}. It was proposed in~\cite{Frandsen2010} that self-interacting ADM may alleviate the solar composition problem but it was shown in \cite{Taoso2010,Cumberbatch2010} that ADM models are incapable of doing so with the cross-sections allowed by direct detection experiments. Recently, there have been improvements of the standard solar model using DM has been achieved with long range interactions~\cite{Lopes:2014aoa} and momentum-dependent cross sections ~\cite{Vincent:2014jia}. Common for the models described above is that they contain only one species of DM particles. In this paper we will study the accumulation of DM in the Sun under the assumption that it is a self-interacting Dirac fermion and that the DM halo is composed of equal amounts of DM and anti-DM particles. This opens up a possibility for an asymmetry in the captured number of DM and anti-DM to occur, either due to different capture cross sections for DM and anti-DM on solar nuclei or due to random fluctuations in the capture process. Moreover, the more abundant species of captured DM can not fully annihilate with its counterpart which allows for large number of captured particles even for an annihilation cross section expected for thermal relic DM. If the size of the population of captured DM of the models considered here can compete with the ADM models, DM of the standard freeze-out scenario may alter solar observables.

The remainder of this paper is organised as follows: In section~\ref{section2}, we present the framework that governs the amount of DM and anti-DM present in the Sun and discuss the behaviour of the DM to anti-DM asymmetry in the limiting cases.
Then, in section~\ref{section3}, we apply our framework to the case of the Sun and in section~\ref{section4} we present our results. Finally, in section~\ref{section5}, we summarise our results and present our conclusions.

\section{Accumulation and the asymmetry} \label{section2}
We model the total number of DM particles in the Sun $N$ and the total number of anti-DM particles in the Sun $\bar N$ using the coupled system of first order differential equations
\begin{eqnarray}
\dot{N}=c+CN+\bar{C}\bar{N}-{\Gamma}N\bar{N} \label{eqn1},
\\
\dot{\bar{N}}=\bar{c}+\bar{C}N+C\bar{N}-{\Gamma}N\bar{N} \label{eqn2}.
\end{eqnarray}
Here, $c$ and $\bar c$ are the capture rates on solar nuclei for DM and for anti-DM, respectively, while $C$ and $\bar{C}$ will depend on the self-capture rates of DM and anti-DM and on the ejection rates of DM and anti-DM that is already trapped by incoming particles from the DM halo. The DM--anti-DM annihilation rate is given by $\Gamma$. In principle, $C$ and $\bar{C}$ also include evaporation effects, but it has been shown that this is negligible for DM particles with a mass $\gtrsim5$~GeV~\cite{Gould1987,Busoni:2013kaa} and we will neglect it in the following discussion.

We define the asymmetry $\Delta$ as the difference between the DM and anti-DM numbers, $\Delta=N-\bar{N}$. It evolves according to
\begin{equation}
\dot{\Delta}=d+D\Delta,
\end{equation}
where we have defined $d=c-\bar{c}$ and $D=C-\bar{C}$. Assuming negligible amounts of both DM and anti-DM at the birth of the Sun, $\Delta(0)=0$, we find
\begin{equation}
\Delta=\int_0^t de^{D(t-\tau)} d\tau.
\end{equation}
An important note is that the absolute value of $\Delta$ also represents the minimum amount of dark particles (DM or anti-DM) present in the Sun. The sum of DM and anti-DM, $N+\bar{N}$, will at any time always be larger or equal to the asymmetry.

Another important note is that there is a geometric limit for the self-capture rates. When $N$ and $\bar{N}$ are large enough, our approach is no longer valid. When DM and anti-DM is trapped inside the Sun, they will accumulate in the center roughly inside a sphere of radius $r_\chi$ (derived in the next section). When $\sigma_{\chi\chi}N+\sigma_{\chi\bar{\chi}}\bar{N}$ is larger than the cross-sectional area of the sphere, ${\pi}r_\chi^2$, every incoming DM particle will scatter off either trapped DM or anti-DM. When the distributions of trapped DM and anti-DM are identical, the fraction of collisions of DM on DM is $f_{\chi\chi}=\sigma_{\chi\chi}N/(\sigma_{\chi\chi}N+\sigma_{\chi\bar{\chi}}\bar{N})$ while the fraction of collisions on anti-DM is given by $f_{\chi\bar{\chi}}=\sigma_{\chi\bar{\chi}}\bar{N}/(\sigma_{\chi\chi}N+\sigma_{\chi\bar{\chi}}\bar{N})$. The same argument holds for incoming anti-DM except the total cross section $\sigma_{\chi\bar{\chi}}N+\sigma_{\chi\chi}\bar{N}$ can not exceed ${\pi}r_\chi^2$. In this case, the fraction of collisions on DM is given by $f_{\chi\bar{\chi}}$ with all $N$ and $\bar{N}$ interchanged. Similarly, the fraction of collisions on anti-DM is given by $f_{\chi\chi}$, again with $N$ and $\bar{N}$ interchanged. When this occurs, the equation for $\Delta$ will fail since eqs. \eqref{eqn1} and \eqref{eqn2} must be corrected in order to take the geometric limit into account.

\subsection{Intrinsic different capture rates}
When the scattering cross section of DM and anti-DM on solar nuclei are different, $c$ and $\bar{c}$ are different and thus $d\neq 0$. We choose $c$ to be larger than $\bar{c}$ (identifying anti-DM as the species with lower capture probability on solar matter). The solution for $\Delta$ is
\begin{equation}
\Delta=\frac{d}{D}(e^{Dt}-1),
\end{equation}
which has three interesting limits:
\begin{equation}
\Delta =
	\begin{cases}
		dt  & \mbox{if } |Dt|\ll1 \\
		-\frac{d}{D} & \mbox{if } |Dt|\gg1, D < 0 \\
		\frac{d}{D}e^{Dt} & \mbox{if } |Dt|\gg1, D > 0
	\end{cases}
\end{equation}
When $Dt$ is small, self-capture is negligible and the asymmetry will be proportional to the difference in the capture rates. When $D<0$, the system eventually stabilizes, since the additional capture of anti-DM by already captured DM balances the difference in the capture rate on normal matter. For $D>0$, DM captures itself at a larger rate than anti-DM. Once this process becomes dominant, it leads to an exponential increase in the amount of DM captured in the Sun.

\subsection{Stochastically induced difference}
When the capture rates are equal ($c=\bar{c}$), the amount of DM might at some point be larger than the amount of anti-DM simply due to random variations in the capture process which can be modelled by adding a white noise signal $\delta_{c}$ to the capture rates
\begin{equation}
c=c_0+\delta_c(t),\;\;\;\langle\delta_{c}(t)\rangle=0,\;\;\;\langle\delta_c(t)\delta_c(\tau)\rangle=s\delta(t-\tau).
\end{equation}
The white noise is normalized such that the expected number of captured DM particles and its variation matches those of a Poisson distribution, i.e., $\langle{n^2}\rangle-{\langle{n}\rangle}^2=\langle{n}\rangle$. We find that
\begin{eqnarray}
\langle{n}\rangle&=&\left\langle\int_0^t c_0+\delta_c(\tau) d\tau\right\rangle=c_0t \\
\langle{n^2}\rangle&=&\left\langle\left(\int_0^t c_0+\delta_c(\tau) d\tau\right)^2\right\rangle={c_0}^2t^2+st
\end{eqnarray}
and hence $s=c_0$. Using the same argument for the capture rate of anti-DM with a white noise signal $\delta_{\bar{c}}$, we find
\begin{equation}
d=\delta_d(t)=\delta_c(t)-\delta_{\bar{c}}(t).
\end{equation}
Since $\delta_c$ and $\delta_{\bar{c}}$ are independent, $\delta_d$ has the properties
\begin{eqnarray}
\langle\delta_{d}(t)\rangle=0,\;\;\;\langle\delta_d(t)\delta_d(\tau)\rangle=2c_0\delta(t-\tau)
\end{eqnarray}
The expectation value of the asymmetry $\Delta$ is zero, which should be expected since the probability of having an over-abundance of DM to anti-DM must be the same as that of having an over-abundance of anti-DM due to symmetry. To estimate the typical magnitude of the asymmetry, we can study the standard deviation of the stochastic variable $\Delta$, given by $\tilde{\Delta}=\sqrt{\langle{\Delta}^2\rangle}$. We find that
\begin{eqnarray} \label{Deltatilde}
{\tilde{\Delta}}^2 = \int_0^t \int_0^t e^{D(2t-\tau-\sigma)}\langle\delta_d(\tau)\delta_d(\sigma)\rangle d{\tau}d{\sigma} =\frac{c_0}{D}(e^{2Dt}-1).
\end{eqnarray}
Thus, the limiting behaviour is similar to the case for intrinsic different capture rates
\begin{equation}
\tilde{\Delta} =
	\begin{cases}
		\sqrt{2c_0t} & \mbox{if } |Dt|\ll1 \\
		\sqrt{-\frac{c_0}{D}} & \mbox{if } |Dt|\gg1, D < 0 \\
		\sqrt{\frac{c_0}{D}}e^{Dt} & \mbox{if } |Dt|\gg1, D > 0
	\end{cases}.
\end{equation}
The major difference is that the short time limit $|Dt| \ll 1$ gives an asymmetry that is expected to grow with the square root of $t$ rather than linearly and that the coefficients are related to $c_0$ and $D$ by a square root.
For small $Dt$, we note that this result is precisely what would be expected from the difference between two Poisson distributions of expectation value $c_0t$ while an equilibrium or an exponential growth occur for strong self-capture.

\section{DM and anti-DM self-capture rates} \label{section3}
For a self-interacting model of DM, the total capture of DM in the Sun will be the sum of a capture rate due to interactions with solar nuclei, a term proportional to the already captured DM and a similar term proportional to the number of captured anti-DM particles. The capture of anti-DM is completely analogous to DM capture although the rates may differ depending on the various scattering cross sections. The specific formulas to compute the capture rates and ejection rates are presented in the appendix but the complexity of the capture rates $C$ and $\bar{C}$ requires some discussion. In what follows, we assume that the time it takes for a captured DM particle to fall into thermal equilibrium in the solar core is negligible.

Generally, the formula for the capture rates of DM by DM and anti-DM as well as the ejection rates are given by:
\begin{equation}
C(\sigma)=\int_0^{R_\odot} 4{\pi}r^2 \int_0^{\infty} \frac{f(u)}{u} w\Omega\;du\;dr.
\end{equation}
The factor $\Omega$ is the rate at which a particle with velocity $w$ will scatter at radius $r$ and contain information on the probability that the incoming particle is captured and whether the target particle is ejected or not.

The rate of capture of an incoming particle without ejecting the target particle is given by
\begin{equation}
C_{s}(\sigma)=\int_0^{R_\odot} 4{\pi}r^2 \int_0^{v_{esc}} \frac{f(u)}{u}{\sigma}n(r)(v_{esc}^2-u^2)\;du\;dr.
\end{equation}
Depending on the collision, the scattering cross section $\sigma$ is either $\sigma_{{\chi}{\chi}}$ or $\sigma_{{\chi}\bar{{\chi}}}$. The velocity of the particle before it falls into the gravitational potential is $u$ and the velocity at radius $r$ is then $w=\sqrt{u^2+v_{esc}^2}$ where $v_{esc}$ is the escape velocity.

The Knudsen number is a measure of the distance DM particles travel on average between collisions and is given by
\begin{equation}
	K=\frac{l(0)}{r_\chi},\;\;\;\;\;\; l(r)=\left( \sum_{\substack{i}} \sigma_i n_i(r) \right)^{-1}.
\end{equation}
The parameter $r_\chi$ is a length scale that describes the size of the distribution. For $K \gg 1$, particles travel a great distance between collisions and an isothermal assumption of the distribution is justified:
\begin{equation}
n_{\rm ISO}(r)=n_{\rm ISO}(0)e^{-\phi(r)/kT},
\end{equation}
where $n_{\rm ISO}(0)$ is the normalization constant, $T$ is the temperature of the distribution and $\phi(r)$ is the gravitational potential energy at radius $r$ from the core which is calculated from
\begin{equation}
\phi(r)=\int_0^r \frac{Gm{_\chi}M(r')}{{r'}^2}\;dr'.
\end{equation}
Here, $M(r)$ is the total mass inside the sphere of radius $r$ around the solar core;
\begin{equation}
M(r)=\int_0^r 4{\pi}\rho(r'){r'}^2\;dr'.
\end{equation}
In the isothermal case, the normalization constant is $n_{\rm ISO}(0)=\pi^{-\frac{3}{2}}r_\chi^{-3}N$ and the length scale $r_\chi$ is defined as $r_\chi^2=3kT_c/2{\pi}G\rho_cm_\chi$ assuming a constant temperature $T(r)=T_{\rm c}$ and density $\rho(r)=\rho_{\rm c}$ with $\rho_{\rm c}$ and $T_{\rm c}$ being the density and temperature in the solar center. Indeed, for a DM particle with mass $m_\chi=5$~GeV, the length scale $r_\chi \sim 0.05\;{\rm R}_{\odot}$ which shows that the vast majority of captured DM particles are concentrated in a very small volume in the center of the Sun. In the case of large scattering cross sections, $K \ll 1$, the particles scatter so often that they will be in local thermal equilibrium with the surrounding nuclei. The distribution for this case was derived in~\cite{Gould1990} which introduces a dependence on the temperature gradient. However, for the cross sections considered in this paper ($\sigma_{SD} \leq 10^{-37}$~cm$^2$), $K \gtrsim 90$ and so we will work with the isothermal distribution.
Defining $\epsilon(r)=n(r)/N$, the distribution can be written $n(r)=\epsilon(r)N$. For anti-DM, the radial distribution is taken to be the same except $N\rightarrow\bar{N}$.

\begin{figure}
\includegraphics[width=0.55\textwidth]{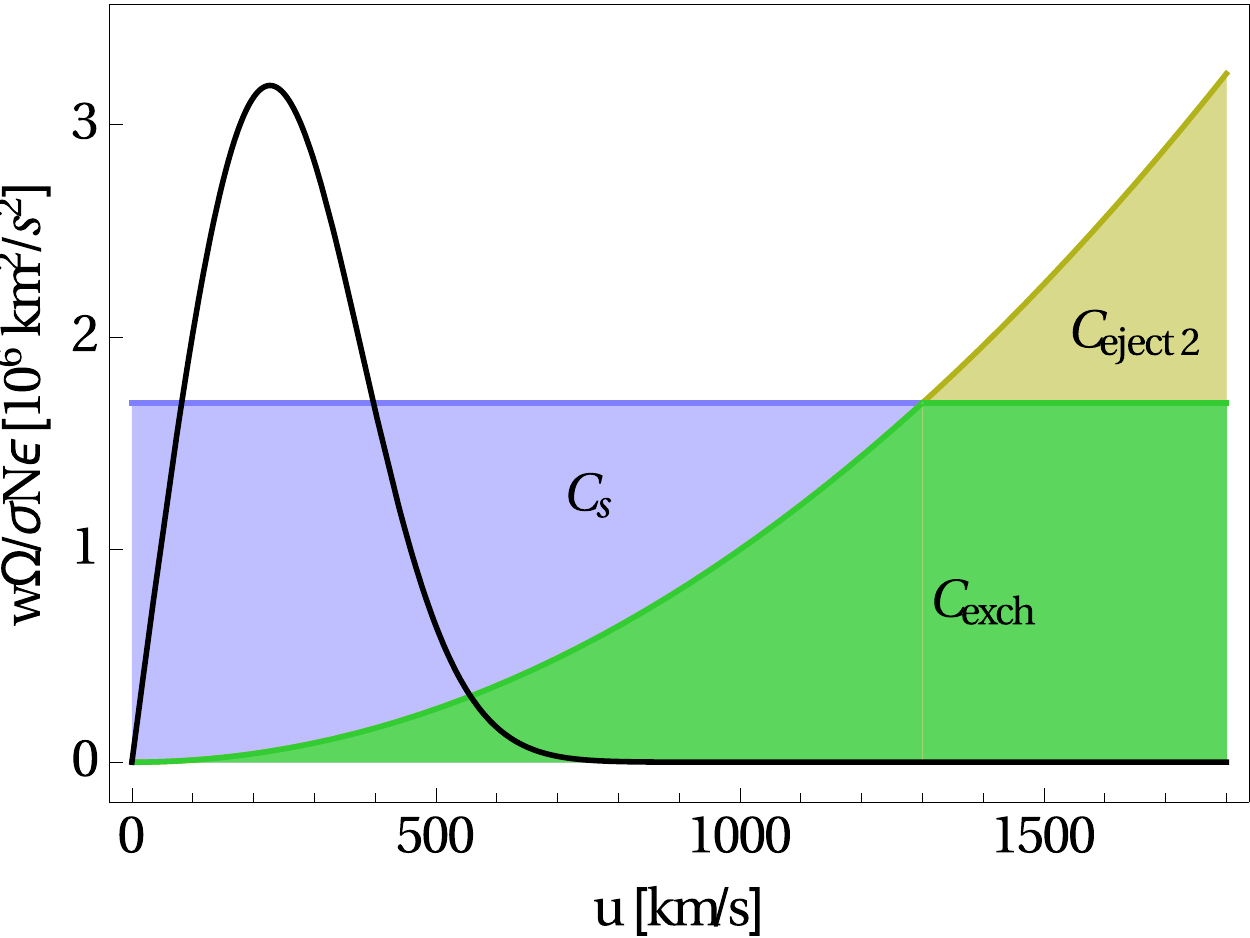}
\caption{The behaviour of the functions $w\Omega$ for different capture scenarios. The horizontal line represents the escape velocity $v_{\rm esc}$ in the center of the Sun, where most of the DM resides. The vertical size of the colored regions represents the quantity which needs to be integrated along with the distribution $f(u)/u$ in order to yield the capture rates. The black curve shows shape of $f(u)/u$ at $v_\odot = 220$~km/s for reference. \label{fig:ejectcapture}}
\end{figure}

The ejection rate of DM captured by in the Sun by collisions with DM or anti-DM from the halo is given by
\begin{equation}
C_{\text{eject}}(\sigma)=\int_0^{R_\odot} 4{\pi}r^2 \int_0^\infty \frac{f(u)}{u}u^2 \sigma N\epsilon(r)\;dudr.
\end{equation}
We must also take into account that while a dark matter particle is being ejected from the Sun, it is also likely that the particle from the halo is captured by the same process. The rate for this exchange occurring is $C_{\rm exch} = C_{\rm eject} - C_{\rm eject\;2}$, where $ C_{\rm eject\;2}$ is the rate of ejections in which both the incoming halo particle and the particle from the Sun are ejected, given by
\begin{equation}
C_{\rm \rm eject\;2} = \int_0^{R_\odot} 4\pi r^2 \int_{v_{\rm esc}}^\infty \frac{f(u)}{u} (u^2-v_{\rm esc}^2) \sigma N\epsilon(r)\; du dr.
\end{equation}
The possible cases depending on velocity for capture and ejection are shown schematically in Fig.~\ref{fig:ejectcapture} along with the velocity distribution of the standard halo model.

As can be seen from this figure, the escape velocity will generally be so large that we expect that the self-capture will be dominant with some contribution from halo particles being captured while ejecting the target particle.

Summarizing this, the capture rates $C$ and $\bar C$ relevant for the evolution of the dark matter and anti-dark matter numbers in the Sun can be written as
\begin{eqnarray}
C &=& C_{s}(\sigma_{{\chi\chi}})-C_{\rm eject}(\sigma_{{\chi}\bar{\chi}})-C_{\rm eject\;2}(\sigma_{{\chi}{\chi}}),
\\
\bar{C}&=&C_{s}(\sigma_{{\chi}\bar{\chi}})+C_{\rm exch}(\sigma_{{\chi}\bar{\chi}}).
\end{eqnarray}
Here, the single self-ejection events do not appear in $C$, as the net change in the DM number in the Sun is zero for these events, but the full ejection induced by the opposite species from the halo must be taken into account as the capture of the halo particle does not compensate for the ejected one. For the capture of DM on anti-DM, the relevant quantities are the capture without ejection and the ejection of the target particle while capturing the halo particle.

The annihilation rate is computed as \cite{Griest1987}
\begin{equation}
\Gamma=\langle{\sigma}v\rangle \int_0^{R_\odot}4{\pi}r^2\epsilon(r)^2\;dr,
\end{equation}
where $\langle{\sigma}v\rangle$ is the thermally averaged annihilation cross section. As long as $R_\odot{\gg}r_\chi$, the upper limit of the integral can be set to $\infty$ rather than $R_\odot$ and the annihilation rate evaluates to
\begin{equation}
\Gamma=\langle{\sigma}v\rangle\frac{1}{(2\pi)^{3/2}r_\chi^3}.
\end{equation}

\section{Results} \label{section4}
In the following, we will make some explicit assumptions in order to estimate the effects described in the previous sections. The velocity distribution $f(u)$ of the halo is assumed to be a standard Maxwell-Boltzmann distribution shifted to the solar frame moving through the halo at $v_\odot=220$ km/s. It can be expressed as \citep{Press1985}
\begin{equation}
f_\chi(u)=n_\chi\frac{u}{\sqrt{\pi}v_{\odot}^2}\left(e^{-\frac{(u-v_{\odot})^2}{v_{\odot}^2}}-e^{-\frac{(u+v_{\odot})^2}{v_{\odot}^2}}\right).
\end{equation}
It is assumed that the DM and anti-DM components in the halo are identical and that the density of each are equal. They will then each have a density $\rho_\chi=\rho_{\bar{\chi}}=0.15$~GeV~cm$^{-3}$, which is equal to half the total local DM density of $0.3$~GeV~cm$^{-3}$ \cite{Bovy2012,Read2014}. The number density of DM and anti-DM is therefore $n_\chi=\rho_\chi/m_\chi$.

The dark matter is assumed to scatter with regular matter with velocity independent spin-independent (SI) and/or spin-dependent (SD) cross sections through effective operators. For the case of SD capture in the Sun, we are interested in the bounds on the SD DM-proton cross section. Limits on these cross sections have been set in various direct detection experiments \cite{LUX2014,XENON100SI,XENON100SD,SuperCDMS2014,Agnese2014,PICASSO2009}. In the DM mass range $10-1000$~GeV, the limit on the SI cross section is $\sigma_{SI}\;{\lesssim}\;10^{-44}$~cm$^2$ \citep{LUX2014}. For smaller DM masses, the limits on the SI cross section weakens significantly. For a $5$~GeV DM particle, $\sigma_{SI}\;{\lesssim}\;{10^{-40}}$~cm$^2$. The limits on the SD cross section in the mass range $10-1000$~GeV is $\sigma_{SD}\;{\lesssim}\;10^{-38}$~cm$^2$ \citep{XENON100SD}. For a $5$~GeV particle, the bound is slightly reduced to $\sigma_{SD}\;{\lesssim}\;10^{-37}$~cm$^2$ \citep{PICASSO2009}.

Limits on the self-interaction of DM comes from astrophysical sources. When galaxy clusters collide, drag forces acting on the gas while the DM passes through unhindered would produce an offset in the mass and gas distribution of the clusters, the size of which can be used to put upper limits on the self-interaction of DM. In~\cite{Randall2008}, one such collision was analysed and and set an upper limit on the self-interacting cross section of $\sigma_{{\chi}{\chi}}/m_{\chi}\lesssim2\cdot10^{-24}$~cm$^2$/GeV.

The relic abundance of DM has been precisely derived from WMAP \citep{Bennett:2012zja} and Planck \citep{Planck:2015xua} experimental data. The thermally averaged annihilation cross section can be related to this relic abundance by solving the Boltzmann equation which is done in e.g., ref~\citep{Jungman1996} and is here taken to be $3\cdot10^{-26}$~cm$^3$/s.

As a model of the Sun, the AGSS09 solar model \cite{Serenelli2009} is chosen. It contains the mass and radial distribution of elements up to Ni. The solar age is taken to be $t_\odot=4.5$ byrs.

The capture rate of a DM particle with mass $m_\chi=5$~GeV is calculated to be at most $10^{27}$~s$^{-1}$ for a SD cross section of $10^{-37}$~cm$^2$ and $2.9\cdot10^{25}$~s$^{-1}$ for a SI cross section of $10^{-40}$~cm$^2$. For a DM particle of mass $10$~GeV, the bounds push SI capture down by four orders of magnitude even though for a fixed SI cross section, the capture rate is only reduced at the percent level. Thus, the SD cross section allows for higher capture rates for all masses in the range 5-1000 GeV.

\begin{figure}
\centering
	\includegraphics[scale=0.6]{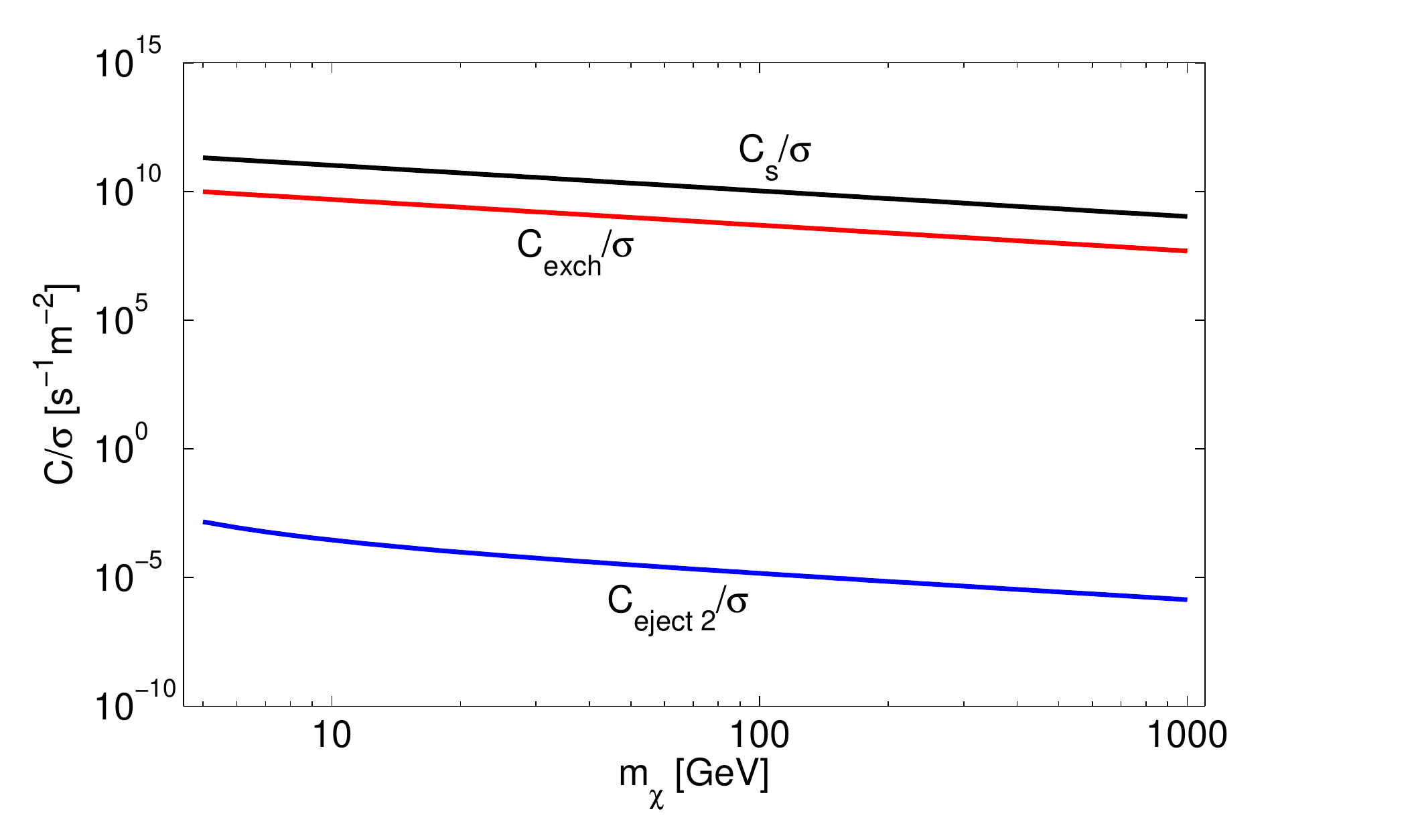}
		\caption{\label{fig:selfcap_ejection} The self-capture rate $C_s/\sigma$ (black line), the exchange rate $C_{\rm exch}/\sigma$ (red line) and the ejection rate $C_{\rm eject\;2}/\sigma$ (blue line) as a function of mass.}
\end{figure}

Fig.~\ref{fig:selfcap_ejection} shows the values of $C_s$, $C_{\rm exch}$ and $C_{\rm eject\;2}$ as a function of DM masses between 5~and~1000~GeV. While $C_{\rm exch}$ is roughly 20 times smaller than $C_s$, $C_{\rm eject\;2}$ is almost 15 orders of magnitude lower. This is not surprising as the escape velocity is very large where the DM resides and particles with a large velocity in the halo are exponentially suppressed (cf.~fig.~\ref{fig:ejectcapture}). In the case of the Sun, $C_{\rm eject\;2}$ may therefore be neglected, since $C_s$ and/or $C_{\rm exch}$ will be completely dominant depending on the relative sizes of $\sigma_{\chi\chi}$ and $\sigma_{{\chi}\bar{{\chi}}}$. The capture rates can now be written as
\begin{eqnarray}
C &\simeq& C_s(\sigma_{\chi\chi}) - C_{\rm exch}(\sigma_{\chi\bar\chi}),
\\
\bar{C} &\simeq& C_s(\sigma_{\chi\bar\chi}) + C_{\rm exch}(\sigma_{\chi\bar\chi}),
\end{eqnarray}
by using that $C_{\rm eject}(\sigma) \simeq C_{\rm exch}(\sigma)$ and one finds that $D$ takes the form
\begin{equation}
D=C-\bar{C}=C_s(\sigma_{\chi\chi})-C_s(\sigma_{{\chi}\bar{\chi}})-2C_{\rm exch}(\sigma_{\chi\bar\chi}).
\end{equation}
Note that even if the scattering cross sections $\sigma_{{\chi}{\chi}}$ and $\sigma_{{\chi}\bar{\chi}}$ are equal, $D$ will be non-zero. This is due to the fact that ejection of the more dominant species occurs at a larger rate.

\subsection{Asymmetric capture and \texorpdfstring{$\Delta$}{Lg}}

\begin{figure}
\centering
	\includegraphics[scale=0.65]{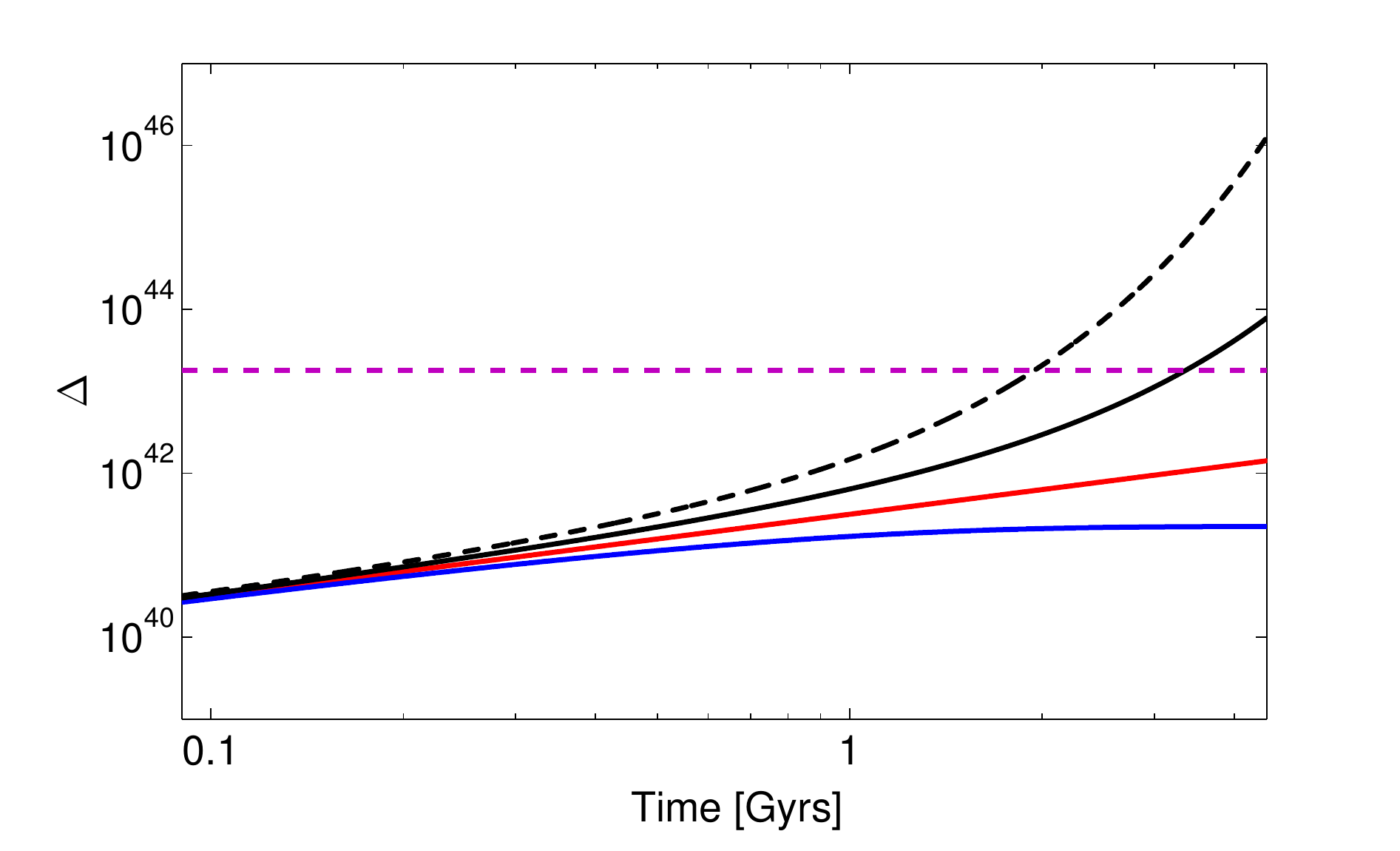}
		\caption{\label{fig:asym1} The evolution of the asymmetry $\Delta$ over time for a $5$~GeV DM particle with a solar capture rate of $c=10^{25}$~s$^{-1}$ and $\bar{c}=0$ and various $\sigma_{\chi\chi}$ and $\sigma_{\chi\bar{\chi}}$. Blue line: $\sigma_{\chi\chi}=0$ and $\sigma_{\chi\bar{\chi}}=3\cdot10^{-24}$~cm$^{2}$, red line: $\sigma_{\chi\chi}=0$ and $\sigma_{\chi\bar{\chi}}=0$, solid black line: $\sigma_{\chi\chi}=3\cdot10^{-24}$~cm$^{2}$ and $\sigma_{\chi\bar{\chi}}=0$, dashed black line: $\sigma_{\chi\chi}=4\cdot10^{-24}$~cm$^{2}$ and $\sigma_{\chi\bar{\chi}}=0$. The purple line shows the geometric self-capture limit for $\sigma_{\chi\chi}=2\cdot10^{-24}$~cm$^{2}$ and $\sigma_{\chi\bar{\chi}}=0$
		}
\end{figure}

Figure \ref{fig:asym1} shows the size of $\Delta$ over time for a DM mass of $5$ GeV, a capture rate of DM on solar nuclei at $c=10^{25}$~s$^{-1}$ and a capture rate of anti-DM on solar nuclei of $\bar{c}=0$~s$^{-1}$, and various different self-scattering cross sections. It can be seen that, when the capture of anti-DM occurs primarily by DM (negative $D$), the asymmetry is smaller than if there would be no self-capture at all or the difference between $\sigma_{\chi\chi}$ and $\sigma_{\chi\bar{\chi}}$ is such that $D$ is small. On the other hand, the exponential growth is apparent when $\sigma_{\chi\chi}$ is larger than $\sigma_{\chi\bar{\chi}}$ as to make $D$ positive. Since $\Delta$ is definitely smaller or equal to $N$, the geometric limit of self-capture has definitely been reached once $\sigma_{\chi\chi}\Delta>{\pi}r_\chi^2$. In the case of fig.~\ref{fig:asym1}, a redefinition of $C$ and $\bar{C}$ would have already been necessary for the two cases with $D>0$.

\begin{figure}
\centering
\begin{minipage}{.5\textwidth}
  \centering
  \includegraphics[scale=0.55]{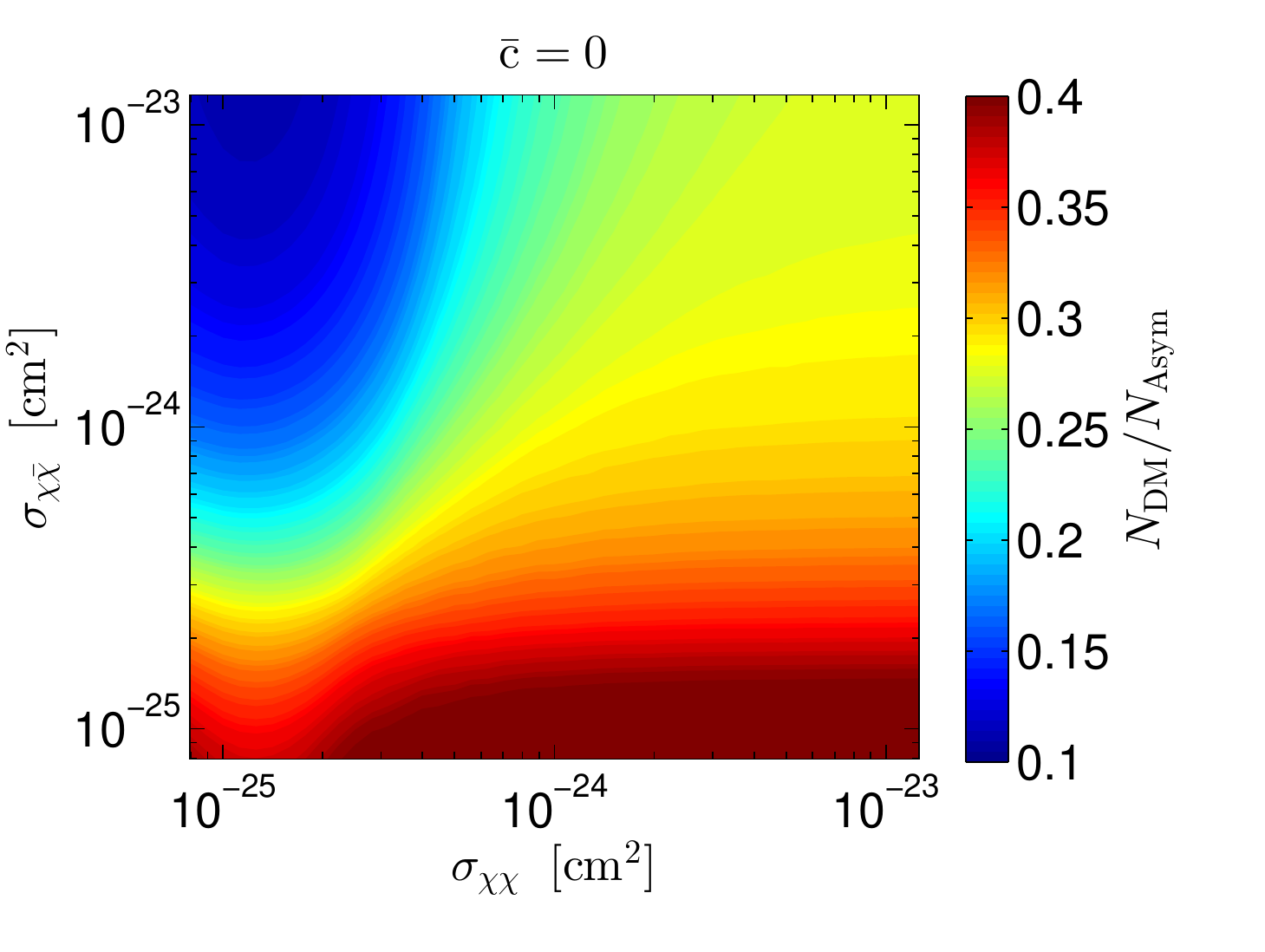}
\end{minipage}%
\begin{minipage}{.5\textwidth}
  \centering
  \includegraphics[scale=0.55]{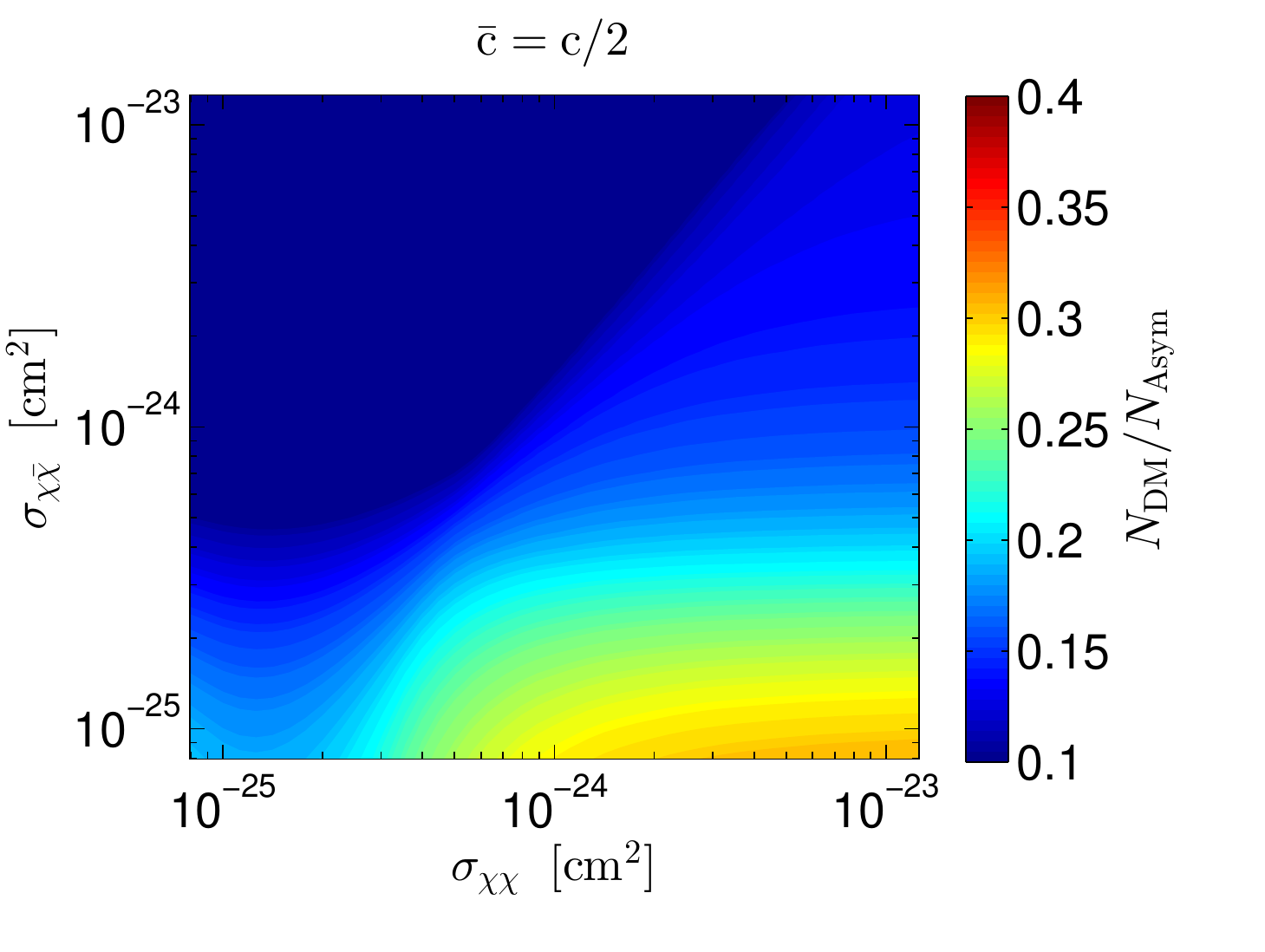}
\end{minipage}
\caption{\label{fig:contourplots} The total number of captured DM and anti-DM particles of mass $m_{\chi}=5$~GeV with a cross section on solar nuclei $\sigma_{\rm SD}=10^{-41}$~cm$^2$ in the $\sigma_{\chi \chi} - \sigma_{\chi \bar{\chi}}$ plane normalized to the number captured by an ADM model with the same DM mass and cross section and twice the halo density ($\rho_{\chi}=0.3$ GeV/cm$^3$). The cross section of anti-DM on solar nuclei is 0 in the left plane and $\sigma_{\rm SD}/2$ in the right.}
\end{figure}

It is of interest to compare the numbers captured in a Dirac DM model to those of an ADM model which is shown in fig.~\ref{fig:contourplots} for two capture rates of anti-DM on solar nuclei. It can be seen that even if $\bar{c}=c/2$, the total number of captured particles in an ADM model, $N_{\rm Asym}$, is at most an order of magnitude larger than the number of captured particles in a Dirac DM model in a large region of the $\sigma_{\chi \chi} - \sigma_{\chi \bar{\chi}}$ plane. Note that in the case that $\bar{c}=0$ and $\sigma_{\chi \bar{\chi}}$ is small, $N_{\rm Asym}$ is slightly larger than half that of the captured numbers in a Dirac model. This is due to the fact that self-capture, $CN$, is limited to a much smaller rate than $c$. The capture of anti-DM is so small that annihilation barely occur and the capture is described approximately by an ADM model in which the background density of DM $\rho_{\chi}$ is halved.

\subsection{Symmetric capture and \texorpdfstring{$\tilde{\Delta}$}{Lg}}

\begin{figure}
\centering
	\includegraphics[scale=0.65]{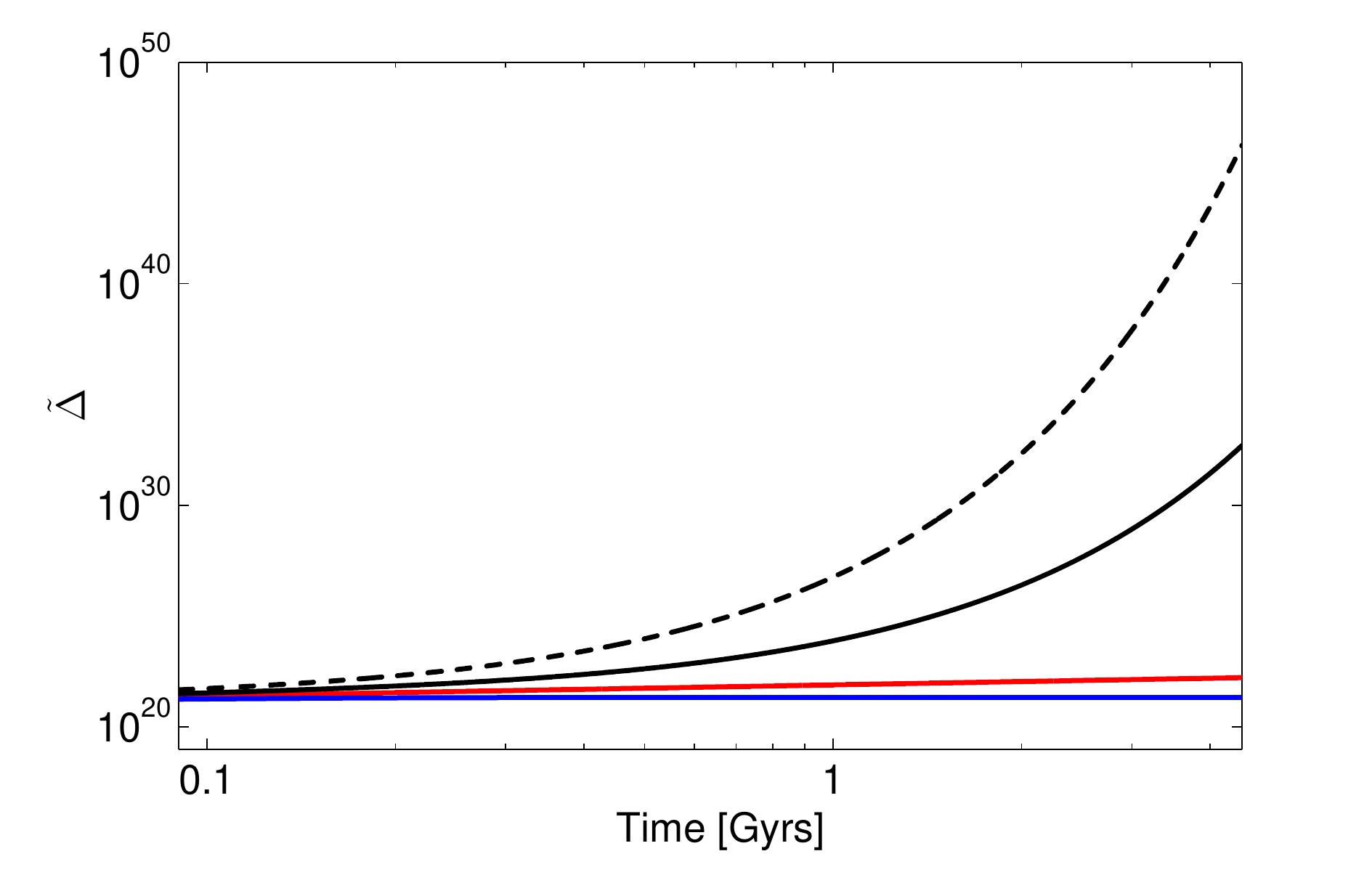}
		\caption{\label{fig:stochasym} The evolution of the stochastic asymmetry $\tilde{\Delta}$ over the lifetime of the Sun for various $\sigma_{\chi\chi}$ and $\sigma_{\chi\bar{\chi}}$ using a $5$ GeV DM mass and a capture rate at $10^{27}$ s$^{-1}$. Blue line:~$\sigma_{\chi\bar{\chi}}=10^{-23}$~cm$^{2}$ and $\sigma_{\chi\chi}=0$, red line: $\sigma_{\chi\chi}=\sigma_{\chi\bar{\chi}}=0$, solid black line: $\sigma_{\chi\chi}=2\cdot10^{-23}$ cm$^{2}$ and $\sigma_{\chi\bar{\chi}}=10^{-23}$ cm$^{2}$, dashed~black~line:~$\sigma_{\chi\chi}=2\cdot10^{-23}$~cm$^{2}$ and $\sigma_{\chi\bar{\chi}}=0$.
		}
\end{figure}

The case of $c=\bar{c}$ implies a simple solution to the steady state of eqs.~\eqref{eqn1} and \eqref{eqn2}. When the capture rates are equal, the symmetry of the equations implies that $N=\bar{N}$. Both the capture of DM and anti-DM will then be given by
\begin{equation}
\dot{N}=c+(C+\bar{C})N-{\Gamma}N^2.
\end{equation}
In the steady state, the capture rate is equal to the annihilation rate so that $\dot{N}=0$ which plugged into the above gives
\begin{equation}
N_\infty=\frac{C+\bar{C}}{2\Gamma}+\frac{\sqrt{(C+\bar{C})^2+4c\Gamma}}{2\Gamma}.
\end{equation}
The total number of captured particles will then be $2N_\infty$.

Fig.~\ref{fig:stochasym} shows the evolution of $\tilde{\Delta}$ over time for various $D$ with a DM mass of $5$ GeV and a capture rate of $10^{27}$ s$^{-1}$. If $Dt$ is small, the stochastic asymmetry at this point in time would be $\tilde{\Delta}=1.7\cdot10^{22}$ while increasing $D$ on the negative side will always lead to a stochastic asymmetry that is smaller. Considering the case of positive $D$, the asymmetry will increase exponentially with $D$. With the same DM mass and solar capture rate but turning on self-capture using $\sigma_{\chi\chi}=2\cdot10^{-23}$ cm$^{2}$ and $\sigma_{\chi\bar{\chi}}=10^{-23}$ cm$^{2}$, the number of particles at which equilibrium occurs is $N_\infty=2.98\cdot10^{41}$ while the size of $\tilde{\Delta}$ is $4.9\cdot10^{32}$. With the same self-scattering cross section and setting $\sigma_{\chi\bar{\chi}}=0$, the stochastic asymmetry becomes 5 orders of magnitude larger than the expected number of particles which slightly decrease to $N_\infty=2.93\cdot10^{41}$. Since $\tilde{\Delta}$ is defined as the standard deviation of $\Delta$, we may in this case expect the asymmetry to be large. However, if the actual asymmetry was of this size, the geometric limit has kicked in and $\tilde{\Delta}$ is no longer given by eq.~\eqref{Deltatilde}. Still, this is an indication that self-capture combined with a stochastically induced asymmetry has lead to a significant accumulation of DM in the Sun. 

\section{Summary and discussion} \label{section5}
In this paper, we have considered the capture of DM in the Sun under the assumption that it is a self-interacting Dirac particle and that the galactic background density consists of equal amounts of DM and anti-DM. This opens up the possibility that a large difference in the captured amount of each type might occur so that the total number of particles in the Sun may continue to grow even though the annihilation cross section is that expected from standard thermal relic DM. The initial asymmetry between the number of captured DM and anti-DM particles can occur either due to different scattering cross sections for DM and anti-DM on solar nuclei or due to stochastic fluctuations in the capture process. Any such asymmetry may then be amplified by self-capture or counter-acted by capture of anti-DM by DM. The size of the asymmetry is independent of the annihilation rate and an analytical expression for its size was derived.

When the capture rates of DM and anti-DM are different ($c\neq\bar{c}$), we have the case of asymmetric capture. If the capture rates of DM by DM is $C$ and the capture rate of DM by anti-DM is $\bar{C}$, we define the difference in these rates as $D=C-\bar{C}$. When $D<0$, the capture of anti-DM is more efficient than DM self-capture which implies that the asymmetry $\Delta$ will at some point find an equilibrium. On the other hand, if $D>0$, DM will capture itself more efficiently so that any initial asymmetry will grow exponentially. The asymmetry is found to become large enough to conclude that the geometric bound on the self-capture needs to be taken into account for a wide range of solar capture rates. This is due to an exponential dependence on the size of the DM on DM and DM on anti-DM capture rates when $D>1/t_\odot$. This occurs for a $5$ GeV DM particle with a capture rate of $10^{25}$ s$^{-1}$ on solar nuclei, corresponding to a spin-dependent cross section of $10^{-39}$ cm$^{2}$, for a $D$ given by $\sigma_{\chi\chi}=2\cdot10^{-24}$~cm$^2$ and $\sigma_{\chi\bar{\chi}}=0$, where $\sigma_{\chi\chi}$ is the DM self-scattering cross section and $\sigma_{\chi\bar{\chi}}$ is the scattering cross section for DM on anti-DM. When the capture rate of anti-DM by DM and the capture rate of DM by DM makes $D$ negative, the asymmetry will always be smaller than in the case when there is no self-capture at all. In any case, the size of the asymmetry implies that the total amount of DM captured may be large. Taking the geometric limit into account and comparing the number of captured DM and anti-DM particles in the Sun by those of an ADM model, it is found that in the region of parameter space where $\sigma_{\chi \chi} \gtrsim \sigma_{\chi \bar{\chi}}$, a sizeable population of captured DM occurs where the numbers are within an order of magnitude than those of an ADM model with similar parameters but a twice as large halo density. Since ADM models may have an impact on solar observables as demonstrated in \cite{Cumberbatch2010}, it is plausible that Dirac models may also have an impact on these. However, in regard to solving the solar composition problem and constraining Dirac DM models, numerical investigations such as those found in \cite{Cumberbatch2010,Taoso2010,Vincent:2014jia} needs to be performed.

When the capture rates on solar nuclei are equal for DM and anti-DM ($c=\bar{c}$), the stochastic asymmetry $\tilde{\Delta}$, which estimates the typical magnitude of the actual asymmetry induced by the stochastic variation of $c$ and $\bar{c}$, is always extremely small in comparison to the total number of particles in the Sun when the self-scattering cross sections $\sigma_{\chi\chi}$ and $\sigma_{\chi\bar{\chi}}$ are such that $D$ is small or negative. However, in the case when $D$ is positive, the exponential dependence of $D$ may bring the stochastic asymmetry to a size several orders of magnitude larger than the expected total number of trapped particles at steady-state with no asymmetry. However, this case is an extreme since the scattering cross section $\sigma_{\chi\chi}$ is taken to be right around the upper bound while $\sigma_{\chi\bar{\chi}}=0$. Increasing $\sigma_{\chi\bar{\chi}}$ to half that of $\sigma_{\chi\chi}$, the stochastic asymmetry is reduced by over 10 orders of magnitude to a negligible level compared to the expected amount for symmetric capture. The window for the asymmetry induced by stochastic variations for the Sun is very small and requires $\sigma_{\chi\chi}\gg\sigma_{\chi\bar{\chi}}$ so it may not be expected that the solar asymmetry is large. However, the capture rates increase proportionally to the background density of DM so that larger self-capture rates may be expected for stars in regions where the background density is larger. Even if the likelihood that the Sun has a negligible asymmetry since the background density is small, stars in such regions may have a $D$ that is several orders of magnitude larger which would increase $\tilde{\Delta}$ significantly thus affecting the evolution of such stars.

In this work, we have neglected the fact that a stochastic capture rate on solar nuclei may imply a stochastic variation of the self-capture rates as well. One such scenario would occur if there are perturbations in the local background density. The investigation of this case is left for future work.

\begin{acknowledgments}
This work was supported by the G\"oran Gustafsson Foundation.
\end{acknowledgments}

\appendix
\section{Capture rates of DM}
The capture of dark matter in celestial bodies is a standard calculation, first done by Press and Spergel \cite{Press1985}, later improved and corrected by Gould \cite{Gould1987a}. Given the velocity distribution of halo dark matter, $f(u)$, in the frame of the Sun where $u$ is the velocity very far away where the gravitational potential of the Sun is negligible. The capture rate is given by
\begin{equation}
c=\int_0^{R_\odot} 4{\pi}r^2 \int_0^{\infty} \frac{f(u)}{u}w\Omega\;du\;dr
\end{equation}
where $w\Omega$ is the rate at which a particle with velocity $w$ at radius $r$ will scatter and lose enough energy to be captured.

\subsection*{Solar element capture}
For the SI cross section, Gould found that
\begin{equation}
w\Omega_i=\sigma_in_i(r)\frac{2E_0}{m_\chi}\frac{\mu_+^2}{\mu}\left[e^{-\frac{m_{\chi}u^2}{2E_0}}-e^{-\frac{\mu}{\mu_+^2}\frac{m_{\chi}w^2}{2E_0}}\right]\theta(\frac{\mu}{\mu_+^2}-\frac{u^2}{w^2})
\end{equation}
while for the SD cross section
\begin{equation}
w\Omega=\sigma_pn_H(r)(w^2-\frac{\mu_+^2}{\mu}u^2)\theta(\frac{\mu}{\mu_+^2}-\frac{u^2}{w^2}).
\end{equation}
$n_i(r)$ is the radial distribution of element $i$ in the Sun. The mass of and scattering cross section on element $i$ is $m_i$ and $\sigma_i$, respectively. The mass of the DM particle is $m_\chi$, $\mu=\frac{m_\chi}{m_i}$ and $\mu_+=\frac{1+\mu}{2}$. The SI scattering cross section scale as
\begin{equation}
\sigma_i=\sigma_pA_i^2\frac{\mu_i^2}{\mu_p^2}
\end{equation}
where $A_i$ is the number of nucleons in element $i$ and $\mu_i$ the reduced mass of element $i$ and the DM. $\sigma_p$ is the proton scattering cross section and $\mu_p$ the reduced mass of the DM and the proton. The more complex formula for SI scattering is due to the form factor which takes the nuclear structure of the target into account for larger energy transfers given by
\begin{equation}
\left|f({\Delta}E)\right|^2 = e^{-{\Delta}E/E_i}
\end{equation}
where
\begin{equation}
E_i = \frac{3\hbar^2}{2M_iR_i^2}, \;\;\;R_i = \left(0.91\left(\frac{M_i}{\text{GeV}}\right)^\frac{1}{3}+0.3\right)\text{fm}
\end{equation}
and $M_i$ is the mass of nuclei $i$. For hydrogen, this form factor is set to unity. The total capture rate for a SI capture rate is then the sum of the capture rate by each individual element. For SD capture, hydrogen is the only element of importance since there is no $A^2$ enhancement of the cross sections on other elements and the fraction of elements with spin is completely negligible compared to hydrogen.

\subsection*{Self-capture}
If DM has a non-zero scattering cross section on other DM and anti-DM particles, it may also be captured by colliding with other DM and anti-DM particles. A derivation of self-capture is given in ref.~\cite{Zentner2009} and we review the result and add three cases for which the target particle is ejected. For self-capture, $\Omega$ is broken down to
\begin{equation}
\Omega={\sigma}n(r)wP_{cap}.
\end{equation}
Here, $\sigma$ is the DM self-scattering cross section, $n(r)$ the radial distribution of already captured DM and $P_{cap}$ the probability that the particle is captured in a collision while not giving the target particle enough energy to escape the Sun. The projectile and target particles are gravitationally unbound when their kinetic energy is greater than $m_{\chi}v_{esc}^2/2$. This means that, for capture of a particle without ejecting the target, the energy transfer ${\Delta}E$ must be in the interval
\begin{equation}
\frac{u^2}{w^2}<\frac{{\Delta}E}{E}<\frac{v_{esc}^2}{w^2}.
\end{equation}
The energy transfer distribution is assumed uniform on the interval which gives $\Omega(r,w)$ as
\begin{equation}
w\Omega={\sigma}n(r)(v_{esc}^2-u^2)\theta(v_{esc}-u)
\end{equation}
and the self-capture rate is given by
\begin{equation}
C_{s}=\int_0^{R_\odot} 4{\pi}r^2 \int_0^{v_{esc}} \frac{f(u)}{u}{\sigma}n(r)(v_{esc}^2-u^2)\;du\;dr
\end{equation}

\subsection*{Ejection}
When the transferred energy in a collision involving DM and anti-DM is greater than $m_{\chi}v_{esc}^2/2$, the particle that is hit will be gravitationally unbound and escape the Sun. This ejection rate is calculated using the same formula as self-capture but with a different $\Omega$. We can divide ejection into two regions, one in which $u<v_{esc}$ and one in which $u>v_{esc}$.

If $u<v_{esc}$ and the incoming particle is trapped after a collision, the target particle may or may not be trapped. If the target is trapped, we have the case of self-capture that is described above. If the target particle is ejected, the transferred energy is in the range
\begin{equation}
\frac{v^2}{w^2}<\frac{{\Delta}E}{E}<1
\end{equation}
and $w\Omega$ is given by
\begin{equation}
w\Omega={\sigma}n(r)u^2\Theta(v_{esc}-u).
\end{equation}

If $u>v_{esc}$, entrapment of the incoming particle will always result in the ejection of the target particle. In the case that the target particle is still trapped, ${\Delta}E$ falls in the interval
\begin{equation}
\frac{u^2}{w^2}<\frac{{\Delta}E}{E}<1.
\end{equation}
The factor $w\Omega$ is then
\begin{equation}
w\Omega={\sigma}n(r)v_{esc}^2\Theta(u-v_{esc}).
\end{equation}
However, if ${\Delta}E$ falls in the interval
\begin{equation}
\frac{v_{esc}^2}{w^2}<\frac{{\Delta}E}{E}<\frac{u^2}{w^2}
\end{equation}
The target particle will be ejected and the incoming particle will still have a velocity that is larger than the escape velocity and thus also escape. For this case, $w\Omega$ is found to be
\begin{equation}
w\Omega={\sigma}n(r)(u^2-v_{esc}^2)\Theta(u-v_{esc}).
\end{equation}

\end{document}